\documentclass[fleqn,10pt,twocolumn]{wlscirep}
\usepackage[T1]{fontenc}
\usepackage{mathtools}
\usepackage{color}
\usepackage{bm}
\usepackage{amssymb}
\usepackage{amsbsy}
\usepackage{breakurl}
\usepackage{dblfloatfix}    % To enable figures at the bottom of page

\title{Simulation platform for pattern recognition based on reservoir computing with memristor networks}

\author[1,2,3,*]{Gouhei Tanaka}
\author[2]{Ryosho Nakane}
\affil[1]{International Research Center for Neurointelligence, The University of Tokyo, Tokyo 113-0033, Japan}
\affil[2]{Department of Electrical Engineering and Information Systems, Graduate School of Engineering, The University of Tokyo, Tokyo 113-8656, Japan}
\affil[3]{Department of Mathematical Informatics, Graduate School of Information Technology and Science, The University of Tokyo, Tokyo 113-8656, Japan}

\affil[*]{gtanaka@g.ecc.u-tokyo.ac.jp}

\begin{abstract}
  Memristive systems and devices are potentially available for implementing reservoir computing (RC) systems applied to pattern recognition. However, the computational ability of memristive RC systems depends on intertwined factors such as system architectures and physical properties of memristive elements, which complicates identifying the key factor for system performance. Here we develop a simulation platform for RC with memristor device networks, which enables testing different system designs for performance improvement. Numerical simulations show that the memristor-network-based RC systems can yield high computational performance comparable to that of state-of-the-art methods in three time series classification tasks. We demonstrate that the excellent and robust computation under device-to-device variability can be achieved by appropriately setting network structures, nonlinearity of memristors, and pre/post-processing, which increases the potential for reliable computation with unreliable component devices. Our results contribute to an establishment of a design guide for memristive reservoirs toward a realization of energy-efficient machine learning hardware.
\end{abstract}

\begin{document}

\flushbottom
\maketitle
\thispagestyle{empty}

%\noindent Please note: Abbreviations should be introduced at the first mention in the main text 'œ no abbreviations lists. Suggested structure of main text (not enforced) is provided below.

\section*{Introduction}

% RC
Machine learning (ML) has been becoming a vital technology in many industries for promoting artificial intelligence (AI) during the last decade. For further penetration and expansion of practical applications based on ML methods, it is often demanded to enhance their computational efficiency by reducing computational time and resources while maintaining desired performance. Reservoir computing (RC) is one of the ML frameworks that can meet such a demand \cite{verstraeten2007experimental,schrauwen2007overview,lukovsevivcius2009reservoir,nakajima2021reservoir}. An RC system is normally composed of an unadaptable dynamic {\it reservoir} for transforming input time series data (or sequential data) into a high-dimensional feature space and a trainable {\it readout} for performing a pattern analysis with a simple learning algorithm. The reservoir needs to be well designed for achieving high computational performance in exchange for the simplicity and speediness of the learning process in the readout. For reservoirs constructed with recurrent neural networks \cite{jaeger2001echo,jaeger2004harnessing,maass2002real}, there are some practical design guides which are mainly helpful in software computation \cite{natschlager2003computer,lukovsevivcius2012practical}.

In the exploration of hardware implementation of the adaptability-free reservoirs, various {\it physical reservoirs} have been developed based on electronics, photonics, spintronics, mechanics, material engineering, robotics, and biology \cite{tanaka2019recent}. A unified viewpoint can be obtained by categorizing the reservoir architectures into the network type \cite{vandoorne2008toward,dockendorf2009liquid,hauser2011towards,kulkarni2012memristor}, the single nonlinear node plus a delayed-feedback type \cite{appeltant2011information,soriano2014delay,larger2017high,dion2018reservoir}, and the excitable continuous medium type \cite{fernando2003pattern,nakane2018reservoir,nakane2021spin}. However, it is still challenging to derive a design guide for each type of physical reservoir. This is partly because of a difficulty in comprehensively understanding how the computational performance of physical RC systems depends on possible influential factors, such as the system architecture, the physical characteristics of system components, and the signal processing method.

% memristive RC
In this study, we tackle the above-mentioned issue by focusing on memristive reservoirs. Memristive systems and devices (or memristors) \cite{chua1971memristor,chua1976memristive} are suitable for constructing physical reservoirs, because their dynamics can show inherent nonlinearity and their resistance, called memristance, can be time-varying based on the history of an applied voltage signal. The nonlinearity and the input history-dependent reaction of memristors are favorable in solving linearly inseparable problems with time series data \cite{lukovsevivcius2009reservoir}. Previous studies have demonstrated the potential of memristive reservoirs for temporal pattern recognition, which can be mainly divided into two types: memristor networks \cite{kulkarni2012memristor,sillin2013theoretical,burger2015computational,lilak2021spoken} and memristor arrays \cite{du2017reservoir,moon2019temporal,zhong2021dynamic,jang2021time}. Memristor networks use complex dynamic behavior of interacting memristors as a reservoir state, whereas memristor arrays leverage a set of nonlinear responses of independent memristors with optional delayed feedback loops. Our target in this study is the network-type memristive reservoirs which are more difficult to design and control compared with the array-type ones. We aim to develop a systematic approach for examining the effects of system components, such as the network architecture and the nonlinear characteristics of memristors, on the computational performance of the memristor-network-based RC systems toward establishing a practical design guide and facilitating their hardware implementation.

% this study
We mathematically formulate a general system of memristor networks and develop a simulation platform for performing temporal pattern classification in the RC framework. In temporal classification tasks, the dataset is given as a set of multiple time series data and the corresponding class labels. Our purpose is to construct a pattern classifier having a high generalization ability by using the memristor-network-based RC system, which can well predict the true class label even for unknown time series data after learning. With our simulation platform, we can examine the effects of various system factors on the computational performance of the memristive RC systems. Numerical experiments are conducted to evaluate the classification performance of the RC systems composed of simple memristor models under different system conditions for three temporal pattern classification tasks: the waveform classification \cite{paquot2012optoelectronic}, the electrocardiogram (ECG) classification \cite{luz2016ecg}, and the spoken digit recognition \cite{rodan2011minimum}. In the waveform classification task with sine and triangular waves, a perfect classification is achieved at the best conditions. In the ECG classification task with normal and abnormal patterns, the classification accuracy reaches a maximum of 86\%. In the spoken digit recognition task with the TI-46 Word corpus, the best classification accuracy is 97.3\%. These classification accuracies are comparable to those obtained by state-of-the-art ML methods. The results suggest that the RC systems with memristor networks are very promising as a building block of next-generation ML and AI hardware.

\section*{Results}

\subsection*{RC systems with memristor networks}

%% fig.1
\begin{figure*}[t]
\begin{center}
\includegraphics[width=0.85\linewidth]{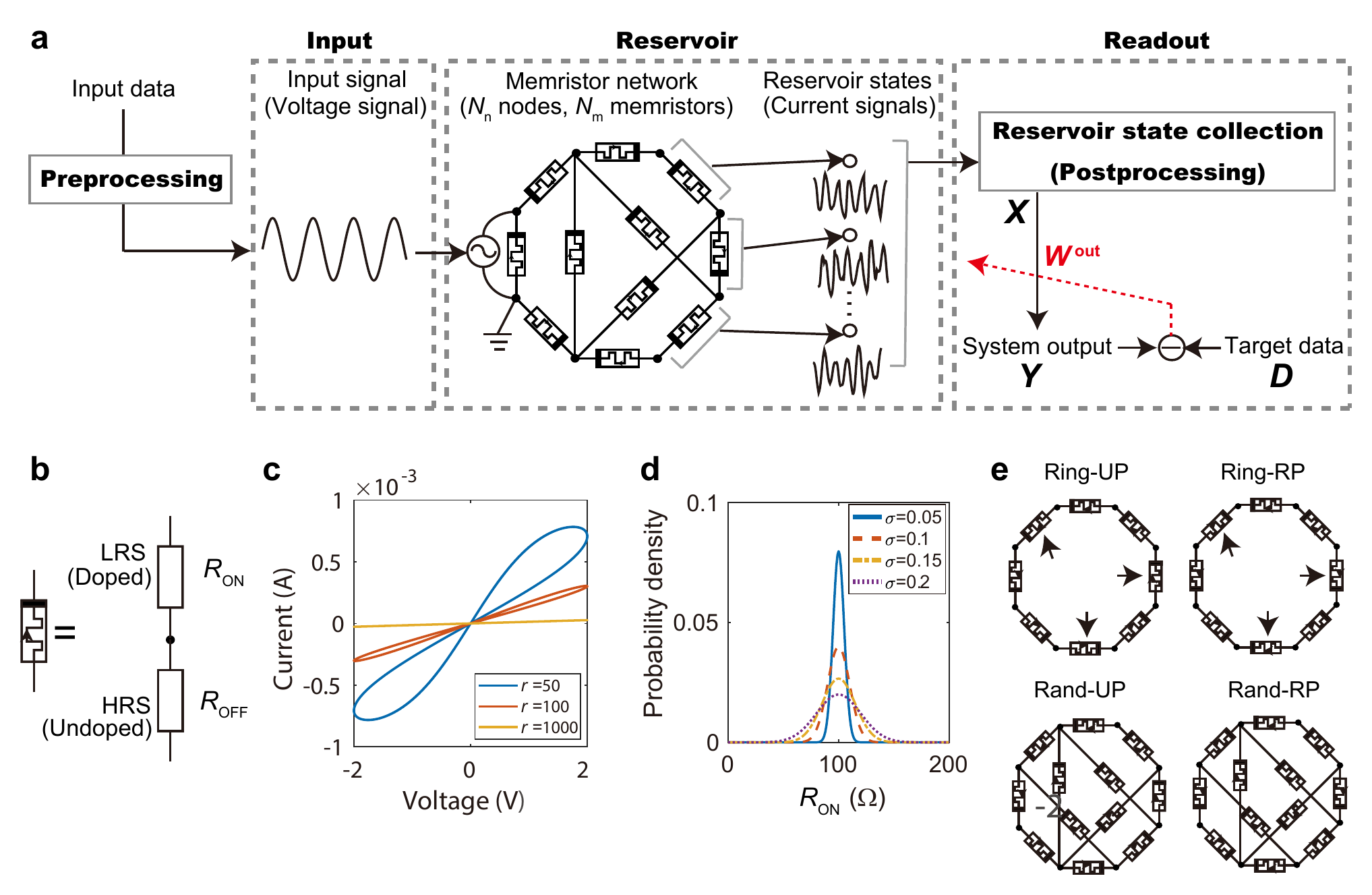}
\caption{{\bf Memristor-network-based RC system.} {\bf a} System architecture composed of preprocessing, input, reservoir, and readout parts. An input time series data is preprocessed and then fed into the memristor network through the voltage source. The memristor network consists of $N_{\rm m}$ memristors, $N_{\rm n}$ circuit nodes, and $N_{\rm i}$ voltage sources. The current signals are measured as a reservoir state and processed in the readout part. The output weight matrix $W^{\rm out}$ is optimized by a linear regression in the training process. {\bf b} The linear drift model of a memristor \cite{strukov2008missing}, which is equivalent to a series of a low resistance state (LRS) with resistance $R_{\rm ON}$ and a high resistance state (HRS) with resistance $R_{\rm OFF} (\gg R_{\rm ON})$. The ratio between the lengths of the LRS and HRS changes in time depending on an applied voltage. {\bf c} Current-voltage ($I$-$V$) characteristics of a single memristor model in response to a sinusoidal voltage signal for different values of $r=R_{\rm OFF}/R_{\rm ON}$. {\bf d} Normal distributions of $R_{\rm ON}$ with mean $\bar{R}_{\rm ON} = 100~\Omega$ and standard deviation $\sigma\bar{R}_{\rm ON}$ for different values of $\sigma$ controlling the degree of the device-to-device variation. {\bf e} Examples of four types of network structures, including a ring with unidirectional polarity (Ring-UP), a ring with random polarity (Ring-RP), a random network with unidirectional polarity (Rand-UP), and a random network with random polarity (Rand-RP). The memristors indicated by the arrows in the Ring-UP and Ring-RP types have opposite polarities. The networks of the Rand-UP and Rand-RP types are generated by randomly adding non-local memristor branches to those of the Ring-UP and Ring-RP types, respectively.}
\label{fig:memristor_network_RC}
\end{center}
\end{figure*}

% system overview
A physical RC system with a memristor network is illustrated in Fig.~\ref{fig:memristor_network_RC}a, which consists of a preprocessing part, an input part, a reservoir part, and a readout part. First, a given time series data is converted to a voltage signal in a preprocessing step, such as appropriate scaling and masking, depending on the type of data. Then, this voltage signal is fed into the voltage source of the memristor-network-based reservoir. A dynamic response of the reservoir to the input signal is obtained as time evolutions of electric currents flowing through the individual memristors. In the readout part, the current signals are converted to a matrix through a collection of reservoir states and an optional postprocessing step, and then, used to produce a system output. Only the output weight matrix $W^{\rm out}$ is trainable, which is optimized by linear regression so as to minimize the error between the system output and the target output. We limit the pre/post-processing methods to linear matrix operations, in order to shed light on the nonlinear transformation effect of the memristive reservoir.

% physical reservoir
We consider a general memristor network consisting of $N_{\rm m}$ memristors, $N_{\rm n}$ circuit nodes, and $N_{\rm i}$ voltage sources (see Fig.~\ref{fig:memristor_network_RC}a with $N_{\rm i}=1$). By regarding the circuit nodes as vertices and the memristor branches as edges, the connectivity of the memristors can be represented as a directional graph, described by an incidence matrix $E_{\rm m} \in \mathbb{R}^{N_{\rm n}\times N_{\rm m}}$. The connectivity of voltage sources can be similarly described by an incidence matrix $E_{\rm i} \in \mathbb{R}^{N_{\rm n}\times N_{\rm i}}$ (see Methods section).

% individual memristors
In this study, we assume that the individual memristors in the reservoir are described by the linear drift model \cite{strukov2008missing}. When an electric voltage is applied to a metal-oxide memristor, the oxygen vacancies (i.e. dopants) drift in the device as charge carriers, and shift the boundary between a doped region with a low resistance and an undoped region with a high resistance. This is simply represented by the linear drift model composed of a series of a low resistance state (LRS) with resistance $R_{\rm ON}$ and a high resistance state (HRS) with resistance $R_{\rm OFF}$ ($\gg R_{\rm ON}$) as illustrated in Fig.~\ref{fig:memristor_network_RC}b (see Methods section for details). With a variation of the length of the LRS, the total memristance changes in time. The polarity of the memristor, related to its directionality, is determined depending on whether the drift of the dopants expands or contracts the LRS \cite{joglekar2009elusive}. For a sinusoidal input voltage, the linear drift model can exhibit a pinched hysteresis loop in the $I$-$V$ curve as shown in Fig.~\ref{fig:memristor_network_RC}c. A variation in the ON/OFF resistance ratio, $r=R_{\rm OFF}/R_{\rm ON}$, changes the nonlinearity of the $I$-$V$ characteristics. When micro/nano-scale memristor devices are fabricated, a device-to-device variation in their electrical properties is inevitable \cite{kuzum2013synaptic,burger2013variation,du2017reservoir,adam2018challenges}. Therefore, $R_{\rm ON}$ and $R_{\rm OFF}$ are assumed to follow a normal distribution of mean $\bar{R}_{\rm ON}$ and $\bar{R}_{\rm OFF}$, respectively (see Methods section). Figure~\ref{fig:memristor_network_RC}d shows a normal distribution of $R_{\rm ON}$ with mean $\bar{R}_{\rm ON}=100~\Omega$ and standard deviation $\sigma \bar{R}_{\rm ON}$ for different values of $\sigma$ controlling the degree of variability. 

Figure~\ref{fig:memristor_network_RC}e illustrates examples of four different types of network structures tested in our numerical simulations, including a ring with unidirectional polarity (Ring-UP), a ring with random polarity (Ring-RP), a random network with unidirectional polarity (Rand-UP), and a random network with random polarity (Rand-RP). These networks are considered to clarify the effect of randomness in network connectivity and polarity. For instance, the difference between Ring-UP and Ring-RP lies in the polarities of the memristors indicated by the arrows. To avoid a disconnected graph, we construct networks of the Rand-UP and Rand-RP types by adding long-range memristor branches to those of the Ring-UP and Rand-RP types, respectively.

%% fig.2
\begin{figure*}[t]
\begin{center}
\includegraphics[width=\linewidth]{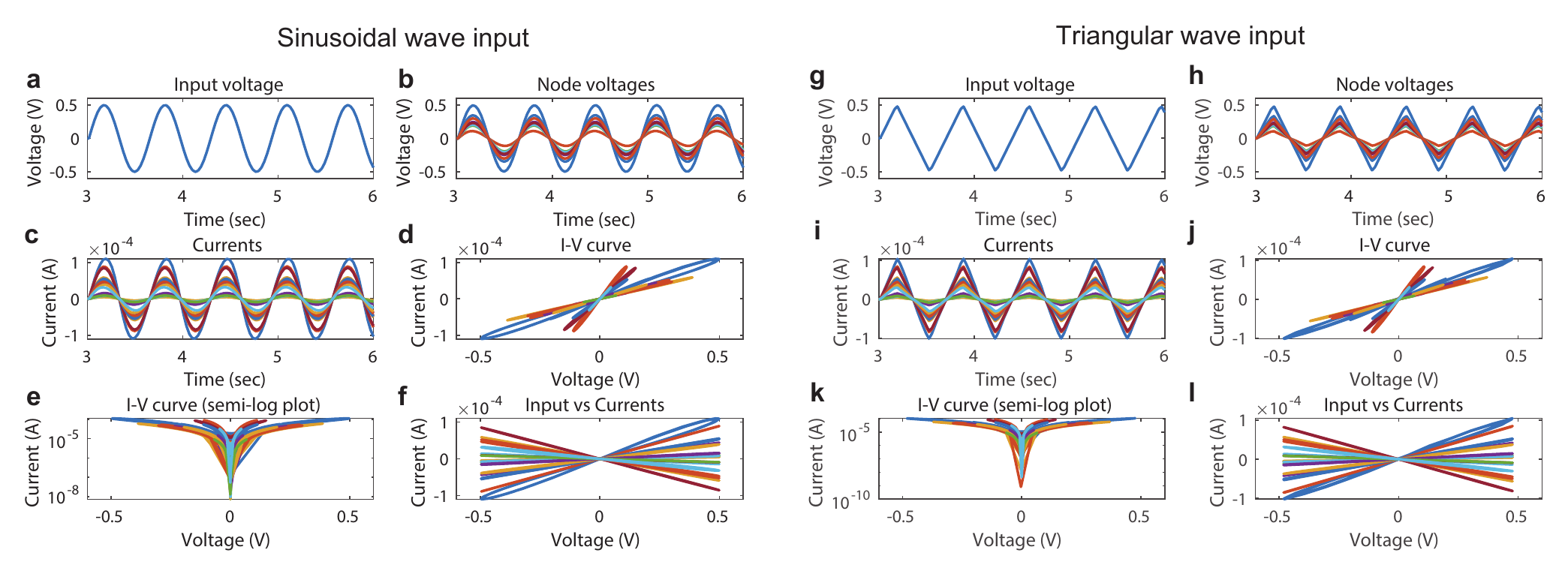}
\caption{{\bf Dynamical behavior of a memristor-network-based reservoir driven by waveform signals.} The network structure is the Rand-UP type, the average ON/OFF resistance ratio is $\bar{r}=50$, and the degree of variability is $\sigma=0.2$. {\bf a} A sine wave input. {\bf b} The nodal voltages. {\bf c} The currents on the memristor branches. {\bf d} The $I$-$V$ curves. {\bf e} The same as {\bf d}, but the absolute current on a semi-log plot. {\bf f} The currents plotted against the nodal voltages. {\bf g}-{\bf l} The same as {\bf a}-{\bf f}, but for a triangular input.}
\label{fig:response_waveform}
\end{center}
\end{figure*}

% circuit equations
We explicitly formulated circuit equations of a general memristor network based on the new modified nodal analysis method \cite{fei2012design} as follows (see Methods section for derivation):
\begin{eqnarray}
  E_{\rm m} W(E_{\rm m}^\top \pmb{\Phi}_{\rm n}(t))E_{\rm m}^\top \frac{{\rm d} \pmb{\Phi}_{\rm n}(t)}{{\rm d}t} + E_{\rm i} \mathbf{j}_{\rm i}(t) &=& \mathbf{0}, \label{eq:kirchhoff_current}\\
\frac{{\rm d} \pmb{\Phi}_{\rm n}(t)}{{\rm d}t} - \mathbf{v}_{\rm n}(t) &=& \mathbf{0}, \label{eq:faraday}\\
E_{\rm i}^\top \mathbf{v}_{\rm n}(t) - \mathbf{v}_{\rm i}(t) &=& \mathbf{0}, \label{eq:kirchhoff_voltage} 
\end{eqnarray}
where $t$ represents continuous time, $\pmb{\Phi}_{\rm n}(t) \in \mathbb{R}^{N_{\rm n}}$ is a vector of nodal magnetic fluxes, $W(\cdot) \in \mathbb{R}^{N_{\rm m}\times N_{\rm m}}$ is a diagonal matrix of memductances (memory conductances) of the memristors as a function of fluxes, $\mathbf{j}_{\rm i}(t) \in \mathbb{R}^{N_{\rm i}}$ is a vector of currents at the voltage sources, $\mathbf{v}_{\rm n}(t) \in \mathbb{R}^{N_{\rm n}}$ is a vector of nodal voltages, $\mathbf{v}_{\rm i}(t) \in \mathbb{R}^{N_{\rm i}}$ is a vector of voltages at the voltage sources. This set of equations with respect to the state variables ($\mathbf{v}_{\rm n}(t)$, $\pmb{\Phi}_{\rm n}(t)$, and $\mathbf{j}_{\rm i}$) are categorized into a differential-algebraic system of equations (DAEs), where differential and algebraic equations are mixed. In general, DAEs are more difficult to solve than ordinary differential equations (ODEs), because the Jacobian matrix (i.e. the first-order derivative) used for numerical integration becomes singular \cite{ascher1998computer}. By setting $E_{\rm m}$, $E_{\rm i}$, $\mathbf{v}_{\rm i}$, and $W$ depending on the reservoir network structure and the memristor models, concrete system equations can be obtained from Eqs.~(\ref{eq:kirchhoff_current})-(\ref{eq:kirchhoff_voltage}) (see Methods section). In our simulation platform on MATLAB \cite{MATLAB2019}, these equations are derived by symbolic computation and solved by using the DAE solver. The dynamic state of the reservoir is represented by time courses of vector $\mathbf{j}_{\rm m}(t) \in \mathbb{R}^{N_{\rm m}}$ of electric currents flowing through the memristors. The reservoir states are used to produce the system output in the readout processing (see Methods section).

%% fig.3
\begin{figure*}[t]
\begin{center}
\includegraphics[width=\linewidth]{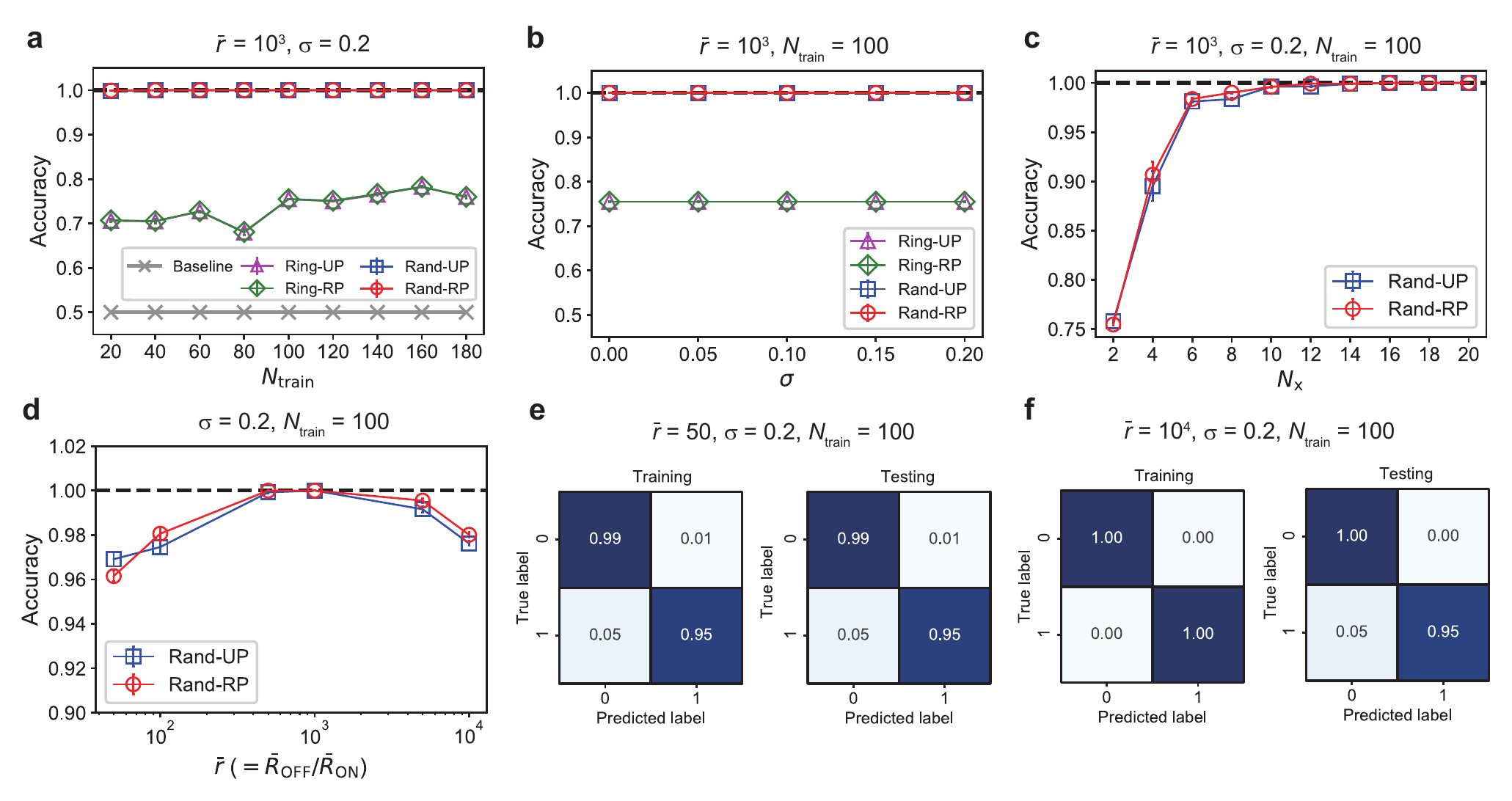}
\caption{{\bf Results of the waveform classification task.} {\bf a} The classification accuracies for the four different network types when $N_{\rm train}$ is varied. {\bf b} The classification accuracies plotted against the variability parameter $\sigma$. {\bf c} The classification accuracies when $N_{\rm x}$ signals out of $N_{\rm m} (=20)$ current signals were used for the readout. {\bf d} The classification accuracies plotted against the average ON/OFF resistance ratio $\bar{r}$. {\bf e} The confusion matrices for $\bar{r}=50$. {\bf f} The confusion matrices for $\bar{r}=10^4$.}
\label{fig:result_waveform}
\end{center}
\end{figure*}

\subsection*{Waveform classification}
% intro
The waveform classification task has often been used as a benchmark task for evaluating the computational performance of physical RC systems \cite{paquot2012optoelectronic,takeda2016photonic,torrejon2017neuromorphic,tanaka2017waveform,zhong2021dynamic}. We consider a two-class classification problem with 100 sine and 100 triangular waves having the same amplitude and different frequencies. The frequency of each data was randomly generated from a uniform distribution in a certain range (see Methods section). Some data were used for training and the other data were for testing. These waveform data were converted to voltage signals and then fed into the memristor-network-based reservoir. The parameter conditions used for numerical experiments are listed in Table~\ref{tab:parameter}.

%% fig.4
\begin{figure*}[t]
\begin{center}
\includegraphics[width=0.9\linewidth]{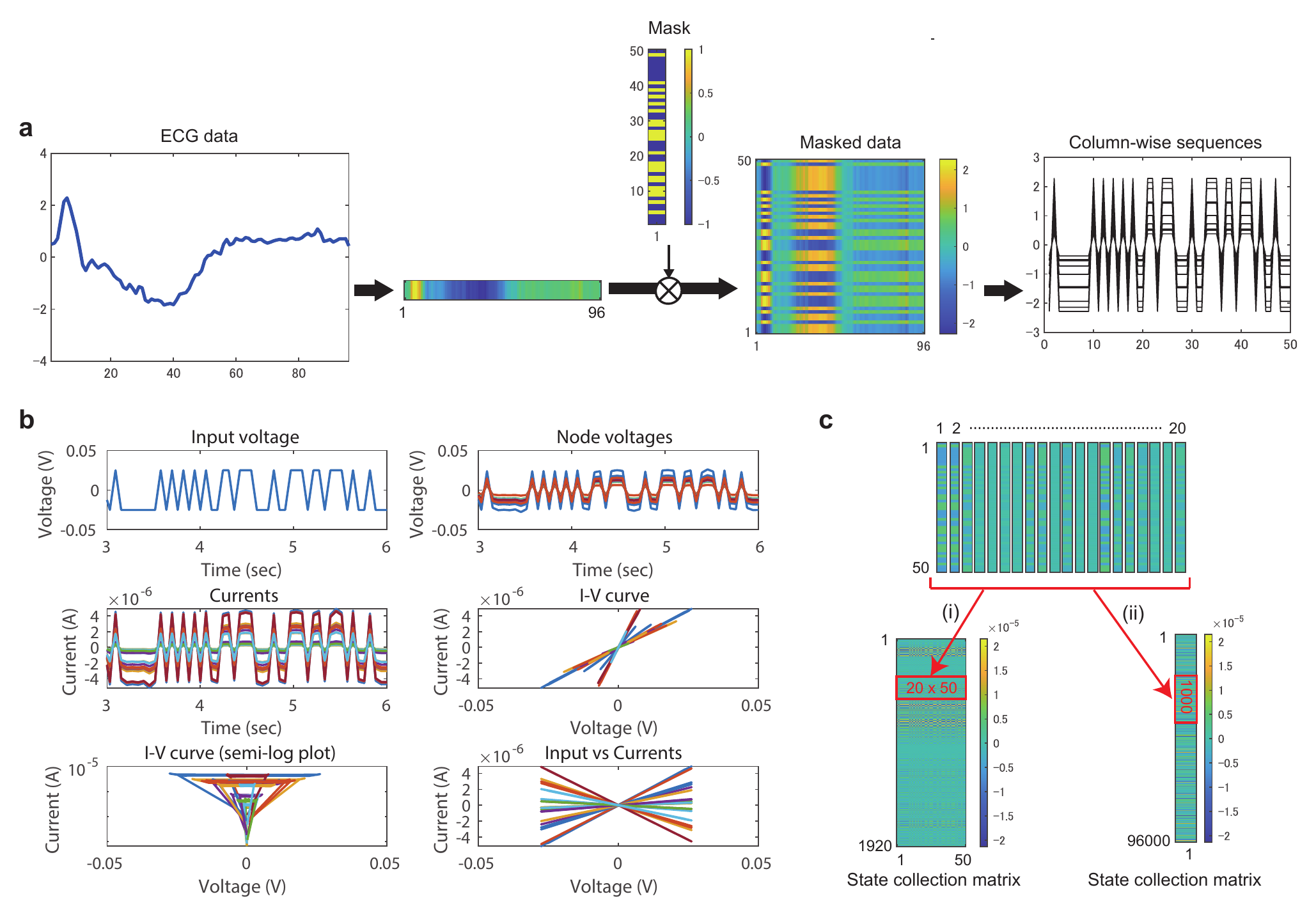}
\caption{{\bf Signal processing for the ECG classification task.} {\bf a} The preprocessing method. An original ECG data is transformed with a binary mask of $+1$ and $-1$ into a masked data. Each of the $L_{\rm data}$~(=96) column-wise sequences is fed into the memristive reservoir. {\bf b} A dynamic response of the memristive reservoir of the Rand-UP type when $\bar{r}=50$ and $\sigma=0.2$. {\bf c} The postprocessing method. The $N_{\rm m}$~(=20) sequences of length $L_{\rm out}$~(=50) are obtained from the reservoir for each input sequence. In Case (i), the reservoir outputs of size $L_{\rm out} \times N_{\rm m}$ are rotated and stacked for all the reservoir inputs to form a state collection matrix. In Case (ii), the reservoir outputs are transformed into a one-dimensional sequence and concatenated for all the reservoir inputs to form a state collection matrix.}
\label{fig:response_ECG200}
\end{center}
\end{figure*}

% nonlinear response
Figure~\ref{fig:response_waveform} demonstrates the responses of a memristor network of the Rand-UP type to sine and triangular input voltage signals when $\bar{r}=\bar{R}_{\rm OFF}/\bar{R}_{\rm ON}=50$ and $\sigma=0.2$. Figures~\ref{fig:response_waveform}a-c show the sine wave input, the nodal voltages, and the currents on the memristor branches, respectively. Due to the random connectivity of the memristors and the device-to-device variation, the individual memristors exhibit different nonlinear $I$-$V$ characteristics as shown in Figs.~\ref{fig:response_waveform}d and e. Figure~\ref{fig:response_waveform}f shows the relationship between the input voltage signal and the branch currents, indicating the input-output transformation realized by the reservoir. The corresponding figures for a triangular wave input are shown in Fig.~\ref{fig:response_waveform}g-l (see Supplementary Figs.~1-5 for other system conditions). The $N_{\rm m}$ current signals were converted to sequences of length $L_{\rm out}=100$ by sampling and then used to form a state collection matrix $X$ for the readout processing (see Methods section).

%% fig.5
\begin{figure*}[t]
\begin{center}
\includegraphics[width=0.8\linewidth]{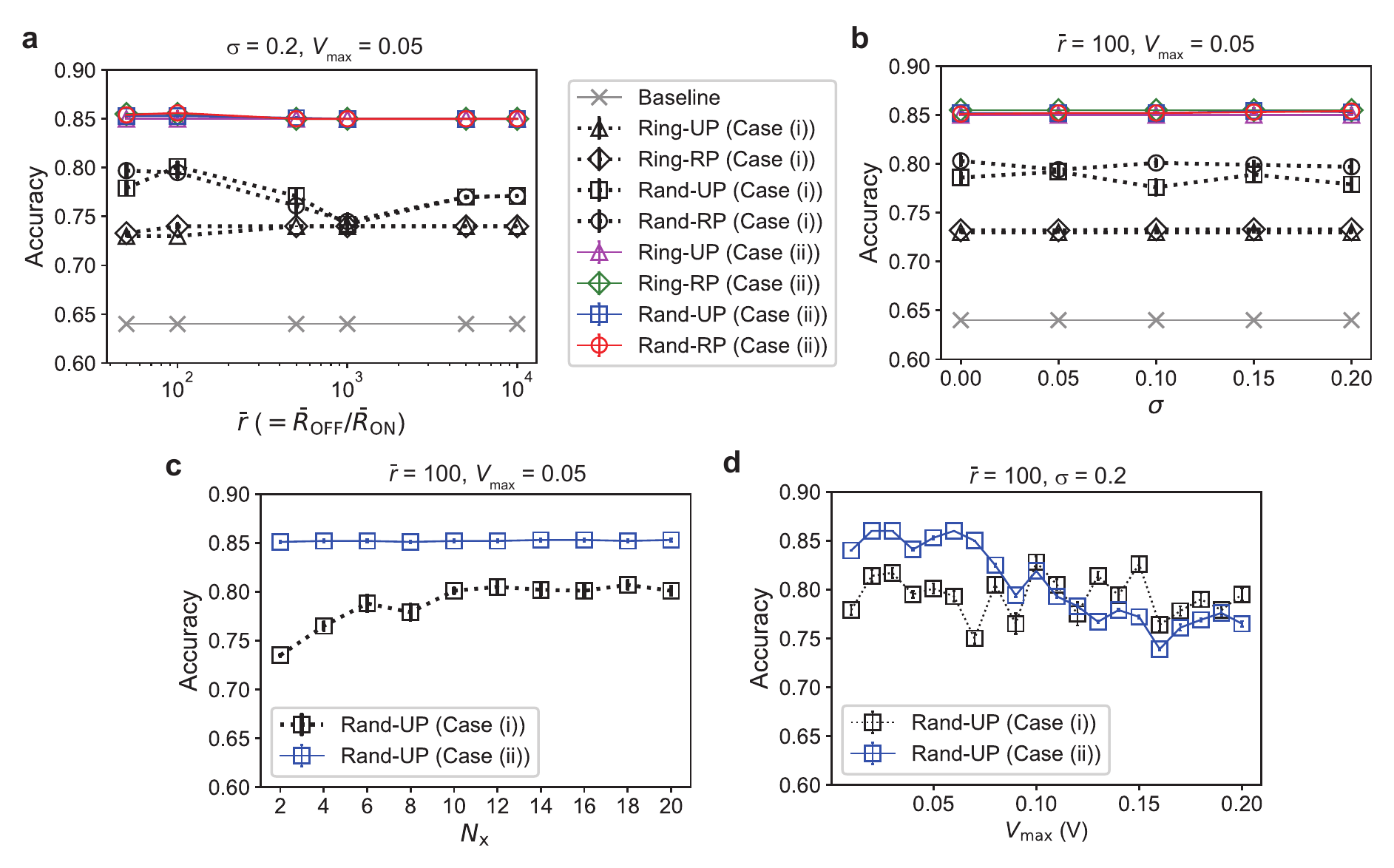}
\caption{{\bf Results of the ECG classification task.} {\bf a} The classification accuracies for a variation of the average ON/OFF ratio $\bar{r}=\bar{R}_{\rm OFF}/\bar{R}_{\rm ON}$. {\bf b} The classification accuracies plotted against the variability parameter $\sigma$. The meanings of the marks are the same as those in {\bf a}. {\bf c} The classification accuracies when $N_x$ signals out of $N_m (=20)$ current signals were used for the readout. {\bf d} The dependence of the accuracy on the maximum input voltage $V_{\rm max}$.}
\label{fig:result_ECG200}
\end{center}
\end{figure*}

% result
Figure~\ref{fig:result_waveform} shows the results of the waveform classification task. We tested the four network structures shown in Fig.~\ref{fig:memristor_network_RC}b: Ring-UP with (magenta triangles), Ring-RP (green diamonds), Rand-UP (blue squares), and Rand-RP (red circles). We evaluated the classification accuracies for 10 random network realizations based on the 10-fold cross validation. Figures~\ref{fig:result_waveform}a-d plot the average value over the 100 trials together with the error bar indicating the standard error. Figure~\ref{fig:result_waveform}a shows the classification accuracies when the number of training data, $N_{\rm train}$, is varied. It is remarkable that the networks of the Rand-UP and Rand-RP types achieve the 100\% accuracy independently of $N_{\rm train}$ due to the rich variety in the reservoir states (see Fig.~2 and Supplementary Fig.~3). The performance of the Ring-UP and Ring-RP types are inferior to that of the Rand-UP and Rand-RP types, but much higher than the baseline accuracy obtained by a linear classifier. This indicates the benefit of the nonlinear signal transformation by the physical reservoir. Figure~\ref{fig:result_waveform}b shows that the classification performance is not sensitive to the change in the variability parameter $\sigma$, suggesting a variability-tolerant property of the memristive RC system. Figure~\ref{fig:result_waveform}c demonstrates that the classification accuracy monotonically increases with the number of signals, $N_{\rm x}$ ($\le N_{\rm m}$), used for the readout. In other words, a higher-dimensional reservoir state yields a better accuracy. The accuracy reaches the perfect level when $N_{\rm x} \ge 14$. Figure~\ref{fig:result_waveform}d shows that the average ON/OFF resistance ratio $\bar{r}$ related to the nonlinearity of the $I$-$V$ characteristics is a key factor influencing the classification performance, which is maximized at the intermediate values around $\bar{r}=5\times 10^2$--$10^3$. Figures~\ref{fig:result_waveform}e and f show the confusion matrices when $\bar{r}=50$ and $\bar{r}=10^4$, respectively. The results imply that the misclassifications are linked to some failures in the training for $\bar{r}=50$ whereas they are caused by over-training for $\bar{r}=10^4$.

\subsection*{ECG classification}
%intro
An electrocardiogram (ECG) is an electric signal associated with heartbeats, which is often used for health checks and cardiac disease detections. In a normal state, the ECG signal shows a repetition of typical waveforms called the QRS complex \cite{pan1985real} due to depolarization and repolarization of the membrane potentials of the cardiac muscle cells. Irregular ECG signals often correspond to abnormal cardiac behavior caused by diseases such as ischemia and myocardial infarction. An ECG-based heartbeat classification task is aimed at separating abnormal ECG signals from normal ones \cite{luz2016ecg,alfaras2019fast}. We used the ECG200 dataset formatted in the UCR Time Series Classification Archive \cite{UCRArchive2018}, which contains a total of 200 samples of ECG segment data, including 100 samples for training and the other 100 samples for testing.

% procedure
Figure~\ref{fig:response_ECG200}a shows a preprocessing step for each ECG data of length $L_{\rm data} (=96)$. The one-dimensional vector representing the original data was transformed into a 2D masked data by using a random mask of size $S_{\rm mask} (=50)$, each element of which is $-1$ or 1. This randomization process corresponds to a multiplication of input data by random input weights in echo state networks \cite{jaeger2001echo,lukovsevivcius2009reservoir}. The masked data was separated into $L_{\rm data}$ column-wise sequences, each of which was converted to a voltage signal and then fed into the memristor-network-based reservoir in the order of the column index sequentially. After the injection of each sequence, we reset the reservoir state. Figure~\ref{fig:response_ECG200}b demonstrates a dynamic response of the reservoir to an input time series. The $N_{\rm m} (=20)$ current signals were transformed into the sequences of length $L_{\rm out} (=50)$ for constructing a state collection matrix $X$ in two possible ways as shown in Fig.~\ref{fig:response_ECG200}c. In Case (i), the transposed matrices of size $N_{\rm m}\times L_{\rm out}$ were stacked vertically for all the inputs to yield a state collection matrix $X \in \mathbb{R}^{N_{\rm m}L_{\rm data} \times L_{\rm out}}$. In Case (ii), the $N_{\rm m}$ sequences were concatenated into a one-dimensional sequence and those for all the inputs were further concatenated to form a state collection matrix (vector) $X\in \mathbb{R}^{N_{\rm m}L_{\rm data}L_{\rm out} \times 1}$.

%% fig.6
\begin{figure*}[t]
\begin{center}
\includegraphics[width=0.8\linewidth]{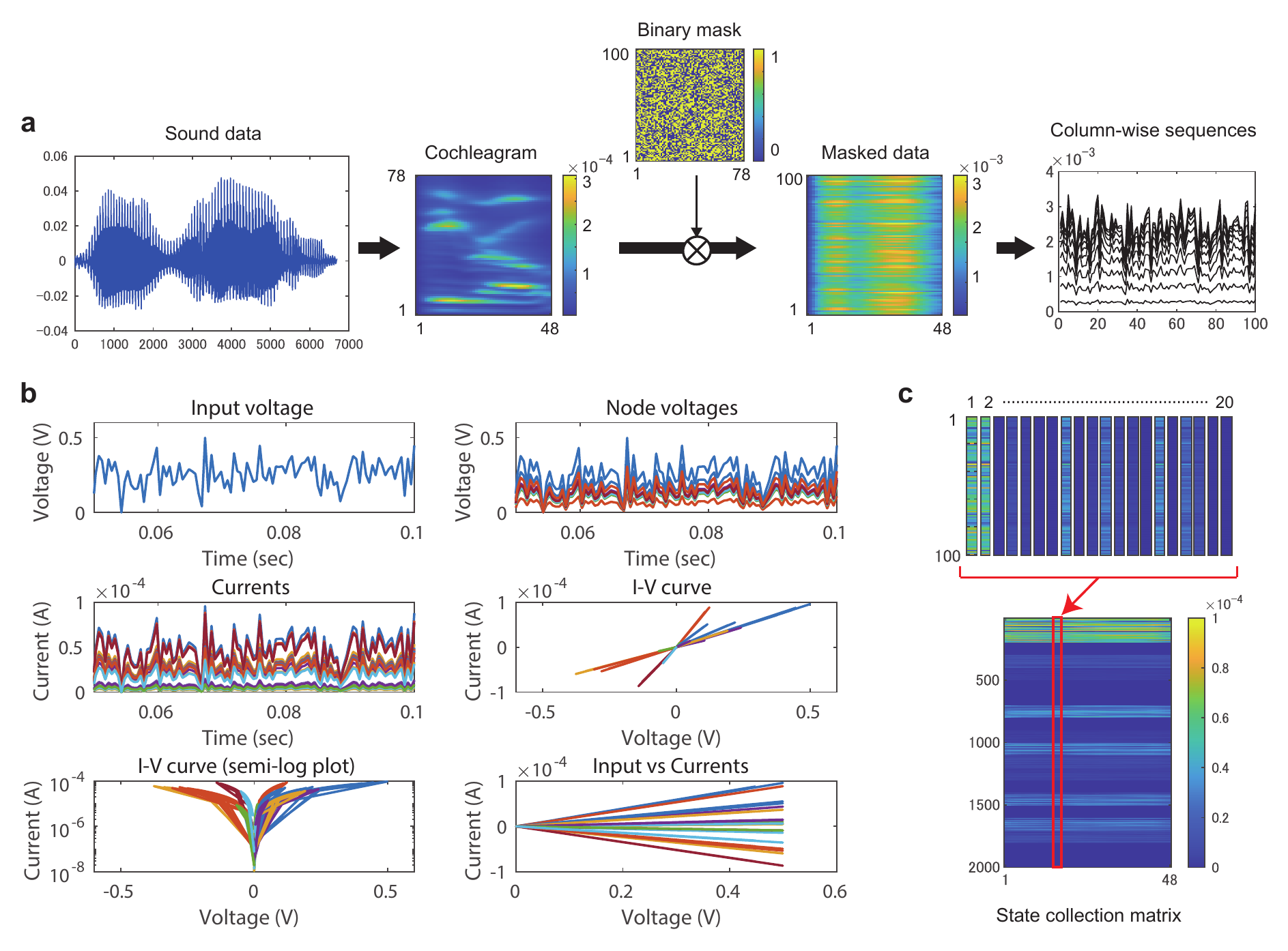}
\caption{{\bf Signal processing for the spoken digit recognition task.} {\bf a} The preprocessing method. A sound data is transformed into a cochleagram of size $N_{\rm f} (=78) \times L_{\rm data} (=48)$ by using the Lyon's passive ear model \cite{lyon1982computational} and then multiplied by a random binary mask of size $S_{\rm mask} (=100) \times N_{\rm f}$. The masked data is separated into $L_{\rm data}$ column-wise sequences. {\bf b} An example of the dynamic response of the memristive reservoir of the Rand-UP type when $\bar{r}=50$ and $\sigma=0.2$. {\bf c} The postprocessing method. The $N_{\rm m}$ current signals are transformed into the sequences of length $L_{\rm out} (=100)$ for each input, which are concatenated into one-dimensional vector of size $N_{\rm m}L_{\rm out}$. The vectors for all the $L_{\rm data}$ inputs are concatenated to make the state collection matrix.}   
\label{fig:response_TI46}
\end{center}
\end{figure*}

% results
Figure~\ref{fig:result_ECG200} shows the results of the ECG classification task. Each plot corresponds to the average accuracy over 10 random network realizations and the error bar indicates the standard error. Figure~\ref{fig:result_ECG200}a plots the classification accuracies for the four types of reservoir structures combined with the postprocessing in Case (i) (monochrome marks) and Case (ii) (colored marks) when $\bar{r}$ is varied. The baseline accuracy (gray crosses) obtained by a linear classifier is 64\%, meaning that all the testing data were classified as normal patterns. We can see that the performance with the postprocessing in Case (ii) are much better than those in Case (i). If the postprocessing method in Case (ii) is employed, the classification performance is kept at a high level against the variation of $\bar{r}$, indicating the robustness against the device-to-device variation as shown in Fig.~\ref{fig:result_ECG200}b. The effect of the number of signals used for the readout processing, $N_{\rm x} (\le N_{\rm m})$, on the performance is shown in Fig.~\ref{fig:result_ECG200}c. By increasing $N_{\rm x}$, the performance is significantly improved in Case (i) but only slightly in Case (ii) (see Supplementary Fig.~6). We observe that the maximum input voltage $V_{\rm max}$ significantly influences the classification performance in Case (ii) as shown in Fig.~\ref{fig:result_ECG200}d. The classification accuracy reaches 86\% when the network structure is the Rand-RP type and $V_{\rm max}=0.02$~V. This performance is comparable to that of the other ML methods, ranging between 77\% and 92\% \cite{ma2019time}. We note that the number of trained weights is 1920 in Case (i) and 96000 in Case (ii). The difference in the number of trainable weights is a potential cause of the gap in the classification performance. These results indicate that the classification performance can largely depend on the postprocessing method even when the same reservoir states are used. It is a future issue to fully understand the influence of the postprocessing on the performance, which is linked with a tradeoff between classification accuracy and computational time for learning.

%% fig.7
\begin{figure*}[t]
\begin{center}
\includegraphics[width=\linewidth]{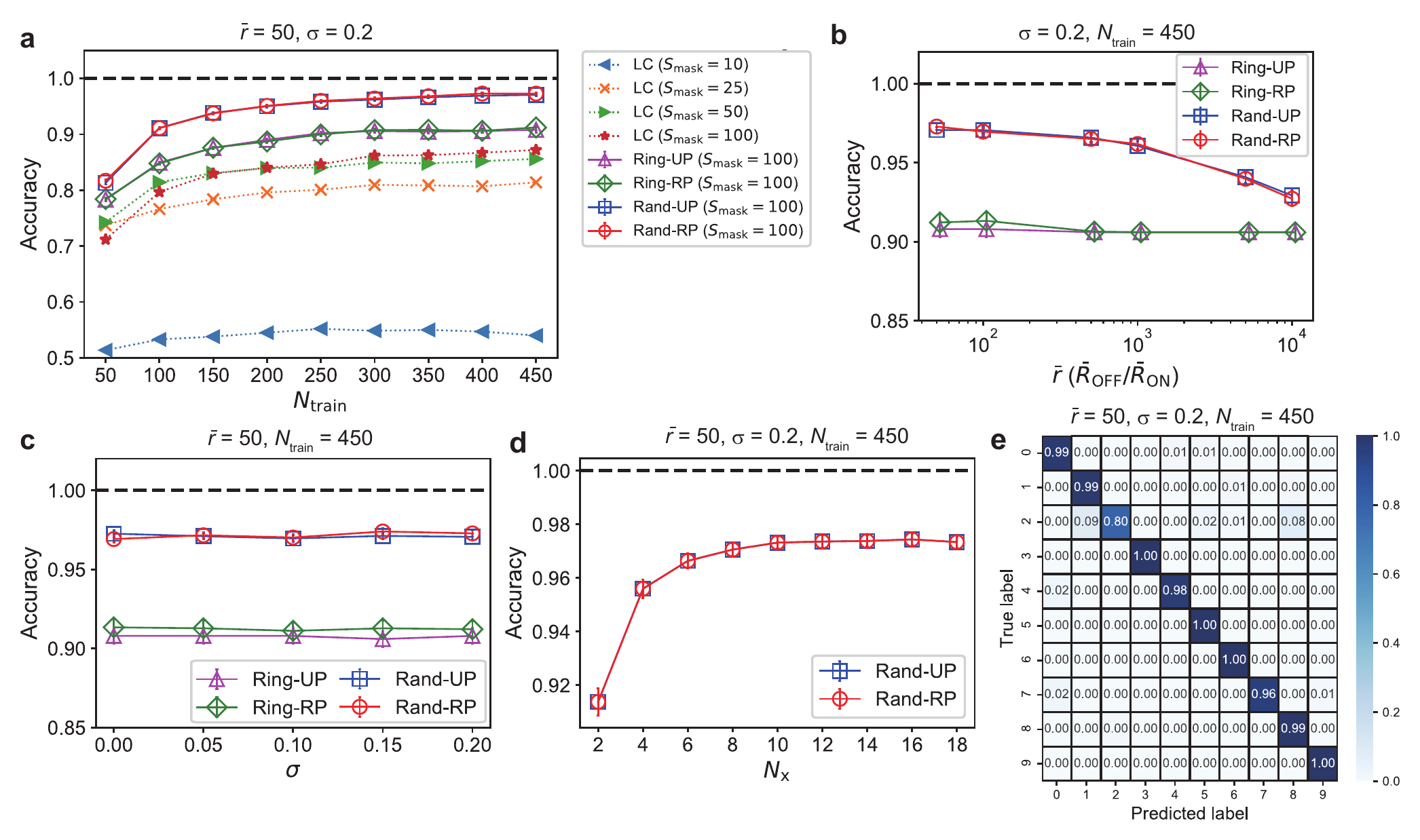}
\caption{{\bf Results of the spoken digit recognition task.} {\bf a} The classification accuracies when the number of training data $N_{\rm train}$ is varied. The results with the linear classifier (LC) include the four cases with $S_{\rm mask}=10, 25, 50, 100$. The results with the memristive RC systems include the four cases with Ring-UP, Ring-RP, Rand-UP, and Rand-RP, where $S_{\rm mask}=100$. {\bf b} The classification accuracies plotted against the average ON/OFF ratio $\bar{r}$. {\bf c} The classification accuracies plotted against the variability parameter $\sigma$. {\bf d} The classification accuracies when $N_{\rm x}$ signals out of $N_{\rm m} (=20)$ current signals were used for the readout. {\bf e} The confusion matrix for testing data.}
\label{fig:result_TI46}
\end{center}
\end{figure*}

\subsection*{Spoken digit recognition}
% intro
The isolated spoken digit recognition has been widely used as a benchmark task for testing the classification ability of RC systems \cite{rodan2011minimum,torrejon2017neuromorphic,tanaka2019recent}. The sound dataset from the NIST TI-46 Word corpus \cite{liberman1993ti} contains 500 sound waveform data corresponding to 10 utterances of 10 digits (``zero'' to ``nine''), spoken by 5 different speakers (ID: 1,2,5,6,7) \cite{verstraeten2005isolated}. The aim of this 10-class classification problem is to correctly predict the true digit from a sound signal.

% preprocessing
Each sound data was converted to a cochleagram by using the Lyon's passive ear model after an elimination of silence periods \cite{lyon1982computational} as shown in Fig.~\ref{fig:response_TI46}a (see Methods section for details). The data length $L_{\rm data}$ ranges from 48 to 102. The cochleagram was transformed into a masked data with a random binary mask of size $S_{\rm mask} \times N_{\rm f}$. Each of the $N_{\rm f}$ column-wise sequences is given to the reservoir after an appropriate scaling. An example of the dynamic behavior of the reservoir is shown in Fig.~\ref{fig:response_TI46}b. The $N_{\rm m}$ (=20) currents from the reservoir are transformed into the sequences of length $L_{\rm out}$ (=100) by sampling. They are concatenated into a one-dimensional vector of size $N_{\rm m}L_{\rm out}$ as shown in Fig.~\ref{fig:response_TI46}c. These column vectors are collected for all the inputs to construct a state collection matrix $X \in \mathbb{R}^{N_{\rm m}L_{\rm out} \times L_{\rm data}}$.

% results
Figure~\ref{fig:result_TI46} shows the results of the spoken digit recognition task. Each plot is the average classification accuracy over 10 random network realizations, evaluated based on the 10-fold cross validation, and the error bar indicates the standard error. Figure~\ref{fig:result_TI46}a demonstrates the classification accuracies when the number of training data, $N_{\rm train}$, is varied. The results obtained by the memristive RC systems with $S_{\rm mask}=100$ are indicated by the colored marks for the four different structure types, i.e. Ring-UP, Ring-RP, Rand-UP, and Rand-RP. For comparison, the performance of linear classifiers (LCs) applied to the masked data are plotted for different mask size $S_{\rm mask}=10, 25, 50, 100$. The classification accuracies for all the tested methods are improved by increasing $N_{\rm train}$. The performance of the LCs increases with the mask size, but peaks out at around $S_{\rm mask}=100$ (see Supplementary Fig.~7). The memristive RC systems produce better accuracies than the LC in the case of $S_{\rm mask}=100$, which highlights the benefit brought about by the nonlinear transformation by the memristive reservoir \cite{araujo2020role}. The network structures of the Rand-UP and Rand-RP types yield better accuracies than those of the Ring-UP and Ring-RP types, owing to the diversity in the reservoir dynamics. Figure~\ref{fig:result_TI46}b shows that a smaller average ON/OFF ratio $\bar{r}$, corresponding to a stronger nonlinearity of $I$-$V$ characteristics on average, leads to better classification performance for the Rand-UP and Rand-RP types when $N_{\rm train}=450$. The best accuracy reaches 97.3\% when the ON/OFF ratio is $\bar{r}=50$ and the network type is Rand-RP, which is comparable to that achieved by other physical RC systems \cite{tanaka2019recent}. For a specific separation between the training and testing datasets among the 10 cases for the cross validation, we obtained 99.8\% accuracy on average over 10 different memristor networks. This is the first report that the memristor networks can yield such high performance in this task. Figure~\ref{fig:result_TI46}c indicates that the classification performance is kept at a high level irrespective of a change in the variability level $\sigma$ of the memristors. Figure~\ref{fig:result_TI46}d demonstrates the results when $N_{\rm x}$ signals out of $N_{\rm m}$ signals were used for the readout process, indicating that a higher dimensional reservoir state contributes more to increasing the classification accuracy and $N_{\rm x}\sim 10$ is enough for obtaining the maximum accuracy. Our analysis on the confusion matrix shown in Fig.~\ref{fig:result_TI46}e reveals that a majority of misclassifications occur when the sound data of digit 2 is incorrectly classified as digit 1 or 8. By overcoming this issue through modifications in the signal processing parts, the performance could be further improved.

\section*{Discussion}
% summary
In this study, we have provided the explicit mathematical formula of the general memristor networks and developed the simulation platform for performing temporal pattern classifications with the memristor-network-based RC systems. The platform enables a considerable search of various system conditions and an identification of the key system component for improving classification performance. The results on the three classification tasks indicate that the randomness of the network connectivity (i.e. Rand-UP and Rand-RP) is favorable for generating diverse nonlinear responses to the input signal and achieving the excellent classification performance compared to the other two (i.e. Ring-UP and Ring-RP). The results have also lead to a new finding that the ON/OFF resistance ratio, controlling the nonlinearity of memristors, can have a large impact on the classification accuracy. Although the best ON/OFF ratios are different depending on the task (see Figs.~\ref{fig:result_waveform}d, \ref{fig:result_ECG200}a, and \ref{fig:result_TI46}b), all of them ($\bar{r}\sim 50$ -- $500$) are within the feasible range reported for real memristor devices \cite{zhao2020reliability}.

% future directions
The memristor network model that we formulated is quite general, and therefore our simulation platform is easily extendable. In this study, we have used at most 20 memristors in the reservoir to test many different system designs while saving the simulation time. Tackling a more complex pattern recognition task with a larger number of memristors is one of the future works. For this purpose, it is an option to inject multiple different input signals into multiple nodes in the memristive reservoir. Any network connectivity of the memristors, including the four types investigated in this study, can be examined conveniently by setting the incidence matrices. We have assumed the ideal linear drift model for the individual memristors. It can be replaced with any other model by deriving or approximating its memductance as a function of the magnetic flux \cite{fei2012design}. In the tested cases, the device-to-device variability following a normal distribution does not cause a degradation in the classification accuracy, implying a reliable computation with heterogeneous units. As far as we checked, the network topology seems to play a decisive role in the input transformation by the reservoir rather than the device-to-device variability. If the shapes and the ranges of the distributions of $R_{\rm ON}$ and $R_{\rm OFF}$ are estimated through measurements with specific devices, they can be easily incorporated into the memristor network model and the simulation platform. We have leveraged appropriate pre/post-processing methods for each task to get high classification accuracies comparable to those obtained by the other ML methods. It would be possible to further improve the classification accuracy by adding more advanced operations in the readout part, but our signal conversion methods limited to linear matrix operations are reasonable for maintaining the merit of low learning cost.

% future goal
  The memristive reservoir of the network type is an attractive option for hardware implementation, because the number of possible network structures, capable of producing different dynamic responses to input signals, can be drastically increased by scaling up the network size. This structural diversity is the major advantage of the network-type reservoir over the array-type and single-node-type reservoirs \cite{tanaka2019recent}. Since only a part of system components are controllable in real memristive devices and materials, it is a significant issue to find better system settings under practical constraints \cite{christensen20212021}. From a scientific perspective, our final goal is to comprehensively understand the relationship between dynamical properties and information processing capacity of memristor networks. The dynamical properties can be investigated through spectral analysis \cite{nakane2021spin} and nonlinear time series analysis \cite{kantz2004nonlinear}. The features related to information processing can be evaluated by computing relevant measures such as memory capacity \cite{jaeger2001short}, kernel quality \cite{legenstein2007edge}, and other capacity scores \cite{dambre2012information}. Our simulation platform can contribute to both these purposes. A target for the future is to integrate numerical and experimental approaches for establishing a design principle of memristive reservoirs, thereby accelerating the development of energy-efficient RC-based ML hardware.

\section*{Methods}

\subsection*{Memristor networks and incidence matrices}
The physical reservoir in this study is a general memristor network consisting of $N_{\rm m}$ memristors, $N_{\rm n}$ circuit nodes, and $N_{\rm i}$ input voltage sources (see Fig.~\ref{fig:memristor_network_RC}a). By regarding the circuit nodes as vertices $v_n$ for $n=1,\ldots,N_{\rm n}$ and the memristor branches as edges $e_m$ for $m=1,\ldots,N_{\rm m}$, a network structure of memristors can be represented as a directional graph with an incidence matrix $E_{\rm m} \in \mathbb{R}^{N_{\rm n}\times N_{\rm m}}$. If a memristor branch $e_m$ ($m=1,\ldots,N_{\rm m}$) connects a starting node $v_k$ and an ending node $v_l$, then the incidence matrix is defined as follows:
\begin{eqnarray}
  E_{\rm m}(n,m) &=&
  \left\{
  \begin{array}{cl}
    -1 & (n=k) \\
    1 & (n=l) \\
    0 & (n \neq k, l)
  \end{array}
  \right. . \label{eq:Em}
\end{eqnarray}
In a similar way, an incidence matrix $E_{\rm i} \in \mathbb{R}^{N_{\rm n}\times N_{\rm i}}$ for the positions of the voltage sources can be defined. If a voltage source $i$ connects a starting node $k$ and an ending node $l$, then
\begin{eqnarray}
  E_{\rm i}(n,i) &=&
  \left\{
  \begin{array}{cl}
    -1 & (n=k) \\
    1 & (n=l) \\
    0 & (n \neq k, l)
  \end{array}
  \right. . \label{eq:Ei}
\end{eqnarray}

\subsection*{Single memristor model}
% linear drift model
The linear drift memristor model \cite{strukov2008missing} was adopted for our simulations. It is assumed that a memristor device consists of a doped region (or LRS) with low resistance $R_{\rm ON}$ and an undoped region (or HRS) with high resistance $R_{\rm OFF}$ (see Fig.~\ref{fig:memristor_network_RC}c). The memristance $M$ at time $t$ is written as follows:
\begin{eqnarray}
M(w(t)) &=& \left( R_{\rm ON}\frac{w(t)}{D}+R_{\rm OFF} \left( 1-\frac{w(t)}{D} \right) \right), \label{eq:M}
\end{eqnarray}
where $D$ and $w(t)$ represent the lengths of the memristor device and the doped region, respectively. When a voltage signal $v(t)$ is applied to the memristor, the current $j(t)$ flowing through the memristor and the time evolution of the internal variable $w(t)$ are described as follows:
\begin{eqnarray}
j(t) &=& v(t)/M(w(t)), \label{eq:i} \\
\frac{{\rm d}w(t)}{{\rm d}t} &=& \mu_v \frac{R_{\rm ON}}{D}j(t), \label{eq:dwdt} 
\end{eqnarray}
where $\mu_v$ represents the average ion mobility. Since the memristance can be written as $M(q)={\rm d}\Phi(q)/{\rm d}q$ using the charge $q$ and the magnetic flux $\Phi$, the memductance $W(q)={\rm d}q(\Phi)/{\rm d}\Phi$ is represented as follows \cite{mcdonald2010analysis}:
\begin{eqnarray}
W(\Phi) &=& (M(w(0))^2-2a\Phi)^{-1/2}, \label{eq:W}
\end{eqnarray}
where $M(w(0))$ is the memristance at the initial condition. The constant parameter $a$ is obtained as follows:
\begin{eqnarray}
a &=& \frac{\mu_vR_{\rm ON}(R_{\rm OFF}-R_{\rm ON})}{D^2}=\mu_v R_{\rm ON} (r-1)/D^2. \label{eq:a}
\end{eqnarray}
where $r$ is the ON/OFF resistance ratio.

\begin{table*}[t]
\begin{center}
\caption{List of system parameters and their values.}
\begin{tabular}{l|l|lll}
\hline
\multicolumn{1}{c}{Parameter} & \multicolumn{1}{c}{Symbol} & \multicolumn{3}{c}{Values} \\
\hline
\hline
\multicolumn{1}{l}{Number of memristors} & \multicolumn{1}{c}{$N_{\rm m}$} & \multicolumn{3}{c}{20} \\
\multicolumn{1}{l}{Number of circuit nodes} & \multicolumn{1}{c}{$N_{\rm n}$} & \multicolumn{3}{c}{10 (Ring-UP/RP), 20 (Rand-UP/RP)} \\
\multicolumn{1}{l}{Number of voltage sources} & \multicolumn{1}{c}{$N_{\rm i}$} & \multicolumn{3}{c}{1} \\
\multicolumn{1}{l}{Device length} & \multicolumn{1}{c}{$D$} & \multicolumn{3}{c}{$10^{-8}$ m \cite{strukov2008missing}}\\
\multicolumn{1}{l}{Average ion mobility} & \multicolumn{1}{c}{$\mu_v$} & \multicolumn{3}{c}{$10^{-14}$ ${\rm m^2s^{-1}V^{-1}}$ \cite{strukov2008missing}}\\
\multicolumn{1}{l}{Mean resistance (Doped)} & \multicolumn{1}{c}{$\bar{R}_{\rm ON}$} & \multicolumn{3}{c}{100 ${\rm \Omega}$ \cite{strukov2008missing}}\\
%\multicolumn{1}{l}{Average ON/OFF resistance ratio} & \multicolumn{1}{c}{$\bar{r}$} & \multicolumn{3}{c}{\{$50$, $10^2$, $5\times 10^2$, $10^3$, $5\times 10^3$, $10^4$\}} \\ 
\multicolumn{1}{l}{Mean resistance (Undoped)} & \multicolumn{1}{c}{$\bar{R}_{\rm OFF}$} & \multicolumn{3}{c}{5k -- 1M~$\Omega$} \\
\multicolumn{1}{l}{Device variation degree} & \multicolumn{1}{c}{$\sigma$} & \multicolumn{3}{c}{ 0 -- 0.2} \\
\hline
\multicolumn{1}{l}{} & \multicolumn{1}{c}{} & \multicolumn{1}{c}{Waveform} & \multicolumn{1}{c}{ECG} & \multicolumn{1}{c}{Spoken digit} \\
\hline
\multicolumn{1}{l}{Number of classes} & \multicolumn{1}{c}{$N_{\rm c}$} & \multicolumn{1}{c}{2} & \multicolumn{1}{c}{2} & \multicolumn{1}{c}{10} \\
\multicolumn{1}{l}{Length of time series data} & \multicolumn{1}{c}{$L_{\rm data}$} & \multicolumn{1}{c}{100} & \multicolumn{1}{c}{96} & \multicolumn{1}{c}{48--102} \\
\multicolumn{1}{l}{Maximum voltage level} & \multicolumn{1}{c}{$V_{\rm max}$} & \multicolumn{1}{c}{0.5~V} & \multicolumn{1}{c}{0.05~V} & \multicolumn{1}{c}{0.5~V} \\
\multicolumn{1}{l}{Mask size} & \multicolumn{1}{c}{$S_{\rm mask}$} & \multicolumn{1}{c}{---} & \multicolumn{1}{c}{50} & \multicolumn{1}{c}{100} \\
\multicolumn{1}{l}{Length of inputs} & \multicolumn{1}{c}{$L_{\rm in}$} & \multicolumn{1}{c}{100} & \multicolumn{1}{c}{50} & \multicolumn{1}{c}{100} \\
\multicolumn{1}{l}{Length of outputs} & \multicolumn{1}{c}{$L_{\rm out}$} & \multicolumn{1}{c}{100} & \multicolumn{1}{c}{50} & \multicolumn{1}{c}{100} \\
\multicolumn{1}{l}{Time length for relaxation} & \multicolumn{1}{c}{$\Delta t_{\rm relax}$} & \multicolumn{1}{c}{3~s} & \multicolumn{1}{c}{3~s} & \multicolumn{1}{c}{0.05~s} \\
\multicolumn{1}{l}{Time length for main signal} & \multicolumn{1}{c}{$\Delta t_{\rm main}$} & \multicolumn{1}{c}{3~s} & \multicolumn{1}{c}{3~s} & \multicolumn{1}{c}{0.05~s} \\
\hline
\end{tabular}
\label{tab:parameter}
\end{center}
\end{table*}

\subsection*{Variability of memristors}
The device-to-device variation can be represented as parameter distributions in the single memristor model. In our study, the resistance $R_{\rm ON}$ of the LRS was generated from a normal distribution with mean $\bar{R}_{\rm ON}$ and standard deviation $\sigma \bar{R}_{\rm on}$. The resistance $R_{\rm OFF}$ of the HRS was generated from a normal distribution with mean $\bar{R}_{\rm OFF} (=\bar{r}\bar{R}_{\rm ON})$ and standard deviation $2\sigma \bar{R}_{\rm OFF}$. It was assumed that $R_{\rm OFF}$ has larger variability than $R_{\rm on}$ \cite{adam2018challenges}. The average ON/OFF resistance ratio $\bar{r}$ controls the nonlinearity of the $I$-$V$ characteristic (see Fig.~\ref{fig:memristor_network_RC}d). The variability parameter $\sigma$ controls the degree of device-to-device variation (see Fig.~\ref{fig:memristor_network_RC}e). The initial condition $w(0)$ was generated from a normal distribution with mean $D/10$ and standard deviation $\sigma D/10$.

\subsection*{Formulation of memristor networks}
A general memristor network can be formulated as Eqs.~(\ref{eq:kirchhoff_current})-(\ref{eq:kirchhoff_voltage}) by using the new modified nodal analysis \cite{fei2012design}. First, the Kirchhoff's current law gives a set of $N_{\rm m}$ differential equations as follows:
\begin{eqnarray}
  E_{\rm m}\mathbf{j}_{\rm m}(t) + E_{\rm i} \mathbf{j}_{\rm i}(t) &=& \mathbf{0}, \label{eq:kirchhoff}
\end{eqnarray}
where $\mathbf{j}_{\rm m}(t) \in \mathbb{R}^{N_{\rm m}}$ is the vector of currents flowing on the $N_{\rm m}$ memristors and $\mathbf{j}_{\rm i}(t) \in \mathbb{R}^{N_{\rm i}}$ is the vector of currents flowing on the $N_{\rm i}$ voltage sources at time $t$. The current vector can be expressed as follows:
\begin{eqnarray}
\mathbf{j}_{\rm m} &=& \frac{\rm d}{{\rm d}t} \mathbf{q}_{\rm m}(E_{\rm m}^\top \pmb{\Phi}_{\rm n}(t)), \label{eq:jm}
\end{eqnarray}
where $\mathbf{q}_{\rm m} \in \mathbb{R}^{N_{\rm m}}$ and $\pmb{\Phi}_{\rm n} \in \mathbb{R}^{N_{\rm n}}$ denote the vectors of the charges and fluxes, respectively. By substituting Eq.~(\ref{eq:jm}) into Eq.~(\ref{eq:kirchhoff}), Eq.~(\ref{eq:kirchhoff_current}) is derived. Using the first-order linearization, Eq.~(\ref{eq:kirchhoff_current}) can be rewritten as follows:
\begin{eqnarray}
E_{\rm m} W_{\rm m} (E_{\rm m}^\top \pmb{\Phi}_{\rm n}(t))E_{\rm m}^\top \frac{{\rm d} \pmb{\Phi}_{\rm n}(t)}{{\rm d}t}+E_{\rm i} \mathbf{j}_{\rm i}(t) = \mathbf{0}, \label{eq:kirchhoff2}
\end{eqnarray}
where $W_{\rm m} \in \mathbb{R}^{N_{\rm m}\times N_{\rm m}}$ denotes the diagonal matrix whose elements are the memductances of the memristors. For the linear drift model, the diagonal elements are given by Eq.~(\ref{eq:W}). Second, the Faraday's law gives Eq.~(\ref{eq:faraday}) which is a set of $N_{\rm n}$ differential equations. Third, the Kirchhoff's voltage law gives Eq.~(\ref{eq:kirchhoff_voltage}) which is a set of $N_{\rm i}$ algebraic equations. As a result, the network of the linear drift memristor models is formulated as a set of differential-algebraic equations (DAEs) consisting of Eq.~(\ref{eq:kirchhoff2}), Eq.~(\ref{eq:faraday}), and Eq.~(\ref{eq:kirchhoff_voltage}). The DAE was numerically solved with the DAE solver ode15i in the software package MATLAB \cite{MATLAB2019}. For a technical reason, a null signal in time period $[0,\Delta t_{\rm relax}]$ for relaxation was added to the main input voltage signal in time period $[\Delta t_{\rm relax}, \Delta t_{\rm relax}+\Delta t_{\rm main}]$ in our simulations. The main input signal with a maximum absolute voltage $V_{\rm max}$ was generated from an input time series of length $L_{\rm in}$ after scaling and interpolation. The parameter values are listed in Table~\ref{tab:parameter}.

\subsection*{Readout}
The vector $\mathbf{j}_{\rm m}(t)$ of $N_{\rm m}$ currents is measured from the memristor network in time period $[\Delta t_{\rm relax}, \Delta t_{\rm relax}+\Delta t_{\rm main}]$. All the $N_{\rm m}$ currents (or partial $N_{\rm x}$ currents) are used to construct a state collection matrix $X$ for the readout through a task-specific postprocessing (see Methods section for specific tasks). Once the state collection matrix $X_k$ corresponding to the $k$-th input time series data is obtained for $k=1,\ldots,N_{\rm train}$, the overall state collection matrix is obtained as follows:
\begin{eqnarray}
  X_{\rm train} &=& [X_1, \ldots, X_k, \ldots, X_{N_{\rm train}}]. \label{eq:X_train}
\end{eqnarray}
Correspondingly, the overall teacher collection matrix is set as follows:
\begin{eqnarray}
  D_{\rm train} &=& [D_1, \ldots, D_k, \ldots, D_{N_{\rm train}}], \label{eq:D_train}
\end{eqnarray}
where $D_k$ is the teacher matrix for the $k$-th input data. The column count of $D_k$ is the same as that of $X_k$ and its row count is the same as the number of classes, $N_{\rm c}$, in the classification task. If the $k$-th input data has label $c_k \in \{1,\ldots,N_{\rm c}\}$, then each column of $D_k$ is given by a one-hot vector $[0,\ldots,1,\ldots,0]^\top \in \mathbb{R}^{N_{\rm c}}$ where only the $c_k$-th element is 1 and the others are 0.

In the training (i.e. learning) phase, the error between the system output $Y_{\rm train} = W^{\rm out} X_{\rm train}$ and the target output $D_{\rm train}$ is minimized by a linear regression. An optimal solution is obtained as follows:
\begin{eqnarray}
  \hat{W}^{\rm out} &=& D_{\rm train}^\dagger X_{\rm train}, \label{eq:regression}
\end{eqnarray}  
where $D_{\rm train}^\dagger$ represents the pseudoinverse matrix of $D_{\rm train}$.

In the testing (i.e. inference) phase, the classification ability of the trained system is evaluated for $N_{\rm test}$ unknown time series data in the testing dataset. For the testing dataset, $X_{\rm test}$ and $D_{\rm test}$ are composed in a similar way to Eq.~(\ref{eq:X_train}) and Eq.~(\ref{eq:D_train}), respectively. The system outputs for the testing data are computed as $Y_{\rm test} = [Y_1,\ldots,Y_k,\ldots, Y_{\rm test}]=\hat{W}^{\rm out} X_{\rm test}$. From each column of $Y_k$, the row index that gives the largest value is recorded. The predicted class $\hat{c}_k$ is determined as the most frequent value in the recorded row indices for all the columns of $Y_k$. By comparing the true class $c_k$ and the predicted class $\hat{c}_k$ for all the $N_{\rm test}$ testing data, the classification accuracy is computed.

\subsection*{Waveform classification}
The sine and triangular waveform data were generated with the following equations:
\begin{eqnarray}
s_{\rm sin}(t) &=& \sin(2\pi f t) \quad\mbox{for}~t\in [0,1], \nonumber \\
s_{\rm tri}(t) &=& \mbox{sawtooth}(2\pi f t+ \pi/2, 1/2) \quad\mbox{for}~t\in [0,1], \nonumber
\end{eqnarray}
where the ``sawtooth'' is a function in MATLAB and the frequency $f$ was randomly drawn from the uniform distribution in $[f_0(1-\delta), f_0(1+\delta)]$. The parameter values were set at $f_0=5$ and $\delta=0.4$. The whole dataset includes 100 sine and 100 triangular waveform data, all of which have length $L_{\rm data} (=100)$. It was separated into $N_{\rm train}$ training data including randomly chosen $N_{\rm train}/2$ data from each class and the remaining $N_{\rm test} (=200-N_{\rm train})$ testing data.

In the postprocessing step, the current signals were converted to sequences of length $L_{\rm out}$ via sampling, and then to positive-valued sequences by taking their absolute values. The state collection matrix $X \in \mathbb{R}^{N_{\rm m}\times L_{\rm out}}$ is constructed by concatenating those sequences for each input time series. The 10-fold cross validation was used to evaluate the classification accuracy.

\subsection*{ECG classification}
This task was performed with the ECG200 dataset from the UCR Timeseries Classification Archive \cite{UCRArchive2018}, which was originally formatted by Olszewski \cite{olszewski2001generalized}. The dataset contains a total of 200 ECG signal data having class labels corresponding to normal and abnormal heartbeats. The dataset consists of $N_{\rm train} (=100)$ training data and $N_{\rm test} (=100)$ testing data, all of which have length $L_{\rm data} (=96)$. The training dataset includes 31 abnormal and 69 normal data. The testing dataset includes 36 abnormal and 64 normal data.

\subsection*{Spoken digit recognition}
This task was performed with the NIST TI-46 Word corpus collected at Texas Instruments in a quiet acoustic enclosure using an Electro-Voice RE-16 Dynamic Cardioid microphone at 12.5kHz sample rate with 12-bit quantization \cite{liberman1993ti}. From each time series data, the main sound signal of length $L_{\rm data}$ were extracted by removing the silence part with the VOICEBOX, a Speech Processing Toolbox for MATLAB \cite{brookes1997voicebox}. The length $L_{\rm data}$ differs depending on the data, ranging from 48 to 102. As shown in Fig.~\ref{fig:response_TI46}a, each sound signal was transformed into a cochleagram based on the Lyon's passive ear model \cite{lyon1982computational} implemented with the Auditory Toolbox in MATLAB \cite{slaney1998auditory}. The cochleagram is represented as a matrix $P \in \mathbb{R}^{N_{\rm f} \times L_{\rm data}}$ where $N_{\rm f}$ is the number of frequency channels. It was converted to a masked data $P_{\rm mask}=QP \in \mathbb{R}^{S_{\rm mask} \times L_{\rm data}}$, where $Q \in \mathbb{R}^{S_{\rm mask} \times N_{\rm f}}$ represents a binary mask of elements 0 and 1. We set $N_{\rm f}=78$. Each of the $L_{\rm data}$ column-wise sequences of length $L_{\rm in} (=S_{\rm mask})$ was converted to a voltage signal by scaling and interpolation, and then fed into the memristive reservoir. 
%$\epsilon_1=0.015$ in the part before the main signals and $\epsilon_2=0.001$ after them

\subsection*{Data availability}
The data that support the findings of this study are available from the authors upon reasonable request. The ECG200 dataset is available from the UCR Timeseries Classification Archive (\url{https://www.cs.ucr.edu/~eamonn/time_series_data_2018/}). The TI-46 word corpus is available from the Linguistic Data Consortium (\url{https://catalog.ldc.upenn.edu/LDC93S9}).

\subsection*{Code availability}
A MATLAB code for simulating memristor networks and reproducing some results will be made publicly available upon publication in the following link: [\url{https://github.com/GTANAKA-LAB/Memristor-Network-Reservoir}].

%\bibliography{reference}

\section*{Acknowledgements}
This work was partially based on results obtained from a project, JPNP16007, commissioned by the New Energy and Industrial Technology Development Organization (NEDO) (GT and RN), and partially supported by JSPS KAKENHI 20K11882 (GT), Project of Intelligent Mobility Society Design, Social Cooperation Program, UTokyo (GT), and AI Center, UTokyo (GT).

\section*{Author contributions statement}
G.T. conceived this study. G.T. and R.N. devised the methods. G.T. conducted numerical experiments and analyses. All authors wrote and approved the manuscript.

\section*{Competing interests}
The authors declare no conflict of interest.

\end{document}